\documentclass[prl,twocolumn,aps,superscriptaddress]{revtex4-1}

\usepackage{bm}%
\usepackage{float}
\expandafter\ifx\csname package@font\endcsname\relax\else
 \expandafter\expandafter
 \expandafter\usepackage
 \expandafter\expandafter
 \expandafter{\csname package@font\endcsname}%
\fi

\usepackage{array}
\usepackage{amsmath,amssymb}
\usepackage{MnSymbol}
\usepackage{tikz-feynman,contour} 
\usetikzlibrary{shapes,arrows,positioning,automata,backgrounds,calc,er,patterns}
\usepackage{feynmf}

\usepackage{graphicx}
\usepackage{mathrsfs}
\usepackage{bm}
\usepackage{mathtools}
\usepackage{epstopdf}
\usepackage{hyperref}

\hypersetup{%
   pdfpagemode=None, 
   pdfstartpage=1,
   pdfmenubar=true,
   pdftoolbar=true,
   colorlinks = true,
   linkcolor=blue,
   citecolor=blue,
   urlcolor=blue,
   bookmarksopen=false
 }
 \usepackage{amsmath}
\usepackage{amsfonts}
\usepackage{mathtools}
\usepackage{graphicx}
\usepackage{fixme}
\usepackage{color}

\begin{document} 

\title{Mediated interactions and photon bound states in an exciton-polariton mixture}




\author{A. Camacho-Guardian} 
\affiliation{Department of Physics and Astronomy, Aarhus University, Ny Munkegade, 8000 Aarhus C, Denmark}
\author{M. A. Bastarrachea-Magnani} 
\affiliation{Department of Physics and Astronomy, Aarhus University, Ny Munkegade, 8000 Aarhus C, Denmark}
\author{G. M. Bruun}
\affiliation{Department of Physics and Astronomy, Aarhus University, Ny Munkegade, 8000 Aarhus C, Denmark}
\affiliation{Shenzhen Institute for Quantum Science and Engineering and Department of Physics, Southern University of Science and Technology, Shenzhen 518055, China}

\begin{abstract}
{
The quest to realise strongly interacting photons remains an outstanding challenge both for fundamental science and for applications. Here, we explore mediated
 photon-photon interactions in a highly imbalanced two-component mixture of exciton-polaritons in a semiconductor microcavity. Using a theory that takes into account 
non-perturbative correlations between the excitons as well as strong light-matter coupling, 
we demonstrate the high tunability of an effective interaction between  quasiparticles formed by  minority component polaritons   interacting 
with a  Bose-Einstein condensate (BEC) of a majority component polaritons. In particular, the interaction, which is mediated by sound modes in the BEC
can be made strong enough to support a bound  state of two quasiparticles.
 Since these quasiparticles consist partly of photons, this in turn corresponds to a dimer state of photons 
propagating through the BEC. This gives rise to a new light transmission line where the dimer wave function is directly mapped onto correlations between the
photons. Our findings open new routes for  highly non-linear optical materials and novel hybrid light-matter quantum systems. 
}
\end{abstract}
\date{\today}
\maketitle
Achieving strong photon-photon interactions provides a pathway to highly non-linear optics with a range of technological applications, and it is therefore 
intensely pursued using a range of different physical platforms. Exciton-polaritons, in short polaritons, are  hybridised states  of light and excitons  in  semiconductors inside 
microcavities that have risen as a promising candidate to realise such strong interactions~\cite{Deng2010,Liew2011,Carusotto2013,Wang2018,Liew2008,Shelykh2010,Amo2010,Gao2012,Nguyen2013,Chang2014,Espinosa2013,Zasedatelev2019,Dreismann2016,Ballarini2013,Sanvitto2016}.  In spite of impressive experimental progress~\cite{Munoz-Matutano2019,Delteil2019}, it has, however, turned out to be difficult to make the 
photon-photon interaction 
sufficiently strong to realise these objectives. Mechanisms to increase the interaction strength include Feshbach resonances~\cite{Takemura2014,Takemura2014b,Takemura2017,Navadeh2019}, dipolar excitons~\cite{Rosenberg2018,Togan2018}, strongly correlated electrons~\cite{Knuppel2019}, and excitons in  Rydberg states~\cite{gu2019}. 
Recently, one has observed the formation of  quasiparticles, coined polaron-polaritons, resulting from the interaction between 
 the excitonic part of the polariton and a surrounding medium consisting either of  excitons in another spin state~\cite{Takemura2014,Navadeh2019} or 
 electrons~\cite{Sidler2016, Emmanuele2019,Tan2020}. An inherent feature of  quasiparticles is that they interact via the exchange of density modulations in the surrounding medium. 
  Such mediated interactions give rise to a range of important many-body phenomena establishing e.g.\ the realm of Landau's liquid theory~\cite{Landau1957,BaymPethick1991book}, leading to 
conventional~\cite{Bardeen1957} and high $T_c$ superconductivity~\cite{Scalapino1995}, and the fundamental interaction in particle 
physics~\cite{Weinberg1995}.

\begin{figure*}
\begin{center}
\includegraphics[width=1.85\columnwidth]{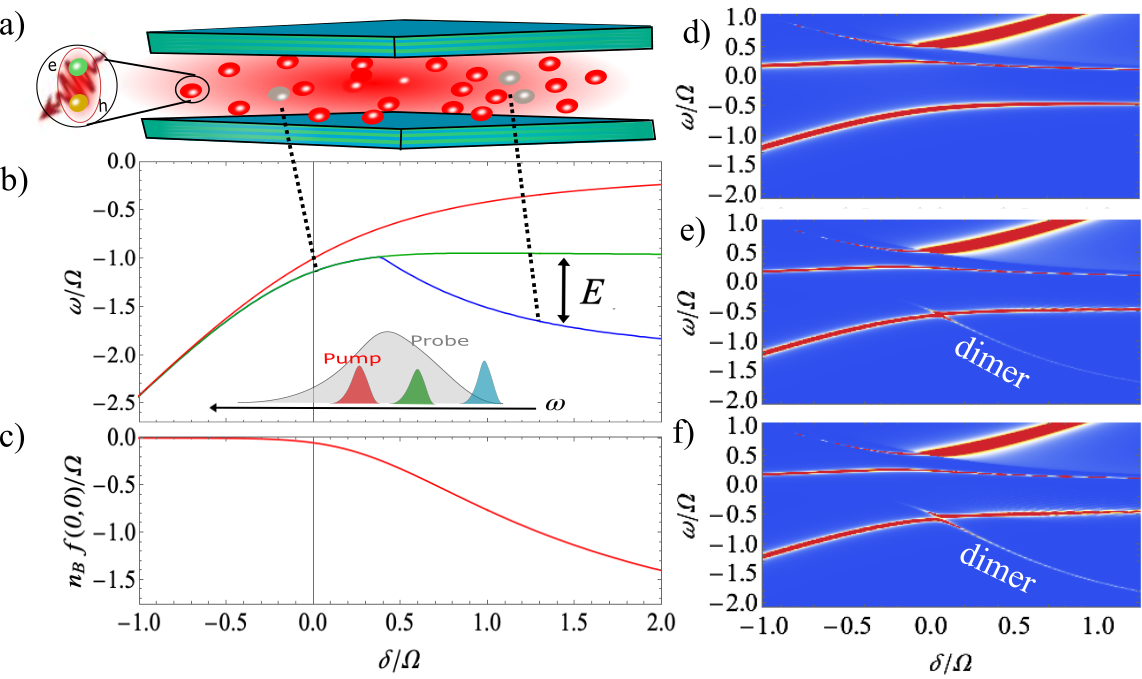}
\end{center}
\caption{(a)  A pump beam creates a BEC of  $\uparrow$ exciton-polaritons (red balls) inside a 2D semiconductor in a microcavity. 
A probe beam creates $\downarrow$ quasiparticles called polaron-polaritons (gray balls), which can bind via an effective interaction mediated by the BEC to form dimer states. (b) The red, green, and blue lines show energy of the $\uparrow$ polaritons,  $\downarrow$ polaron-polaritons, and the $\downarrow\downarrow$ dimers respectively as a function of the detuning $\delta$. (c) The Landau effective interaction between the $\downarrow$ polaron-polaritons. The resulting photon spectral function giving the  transmission of the 
probe beam through the semiconductor is plotted for $\downarrow$  polaron-polariton density $n=0$, $n=0.075n_B$, and (f) $n=0.15n_B$ in d), e), and f). 
This illustrates the emergence of a  distinct  transmission line carried by the dimer states involving two photons propagating through the BEC for $n>0$.  }
\label{OverviewFig}
\end{figure*}

Here, we explore mediated interactions in a highly imbalanced two-component mixture of polaritons created by a pump-probe scheme inside a two-dimensional (2D) semiconductor microcavity as illustrated in Fig.~\ref{OverviewFig}(a). We develop a strong coupling  theory  describing the effective interaction between 
two quasiparticles formed by a minority component polaritons interacting with a surrounding BEC of  majority polaritons. This interaction is shown to be long-range, attractive, and tuneable and as a striking consequence, it supports bound states of two quasiparticles. Since these dimer states partly 
consist of two photons, their propagation through the BEC leads to the 
 emergence of an additional line in the  
light transmission spectrum, see Fig.~\ref{OverviewFig}(a).  
The dimer wave function is moreover shown to be imprinted  on the 
  correlations of the transmitted photons allowing for a direct detection. 

\emph{System.--}
 We consider a  2D mixture of exciton-polaritons in    spin states $\sigma=\uparrow,\downarrow$. The Hamiltonian  is 
\begin{gather}
\hat{H}=\sum_{\mathbf{k}\sigma}
\begin{bmatrix}\hat x_{\mathbf{k}\sigma}^\dagger&\hat c_{\mathbf{k}\sigma}^\dagger\end{bmatrix}
\begin{bmatrix}
\varepsilon_{\mathbf{k}}^{x} & \Omega/2 \\
\Omega/2 & \varepsilon_{\mathbf{k}}^{c}
\end{bmatrix}
\begin{bmatrix}
\hat x_{\mathbf{k}\sigma} \\
\hat c_{\mathbf{k}\sigma}
\end{bmatrix}\nonumber\\
+g_{\uparrow\downarrow}\sum_{\mathbf{q}}\hat\rho_{-\mathbf{q}\uparrow}\hat\rho_{\mathbf{q}\downarrow}
+\frac{g_{\uparrow\uparrow}}2\sum_{\mathbf{q}\sigma}\hat\rho_{-\mathbf{q}\sigma}\hat\rho_{\mathbf{q}\sigma},
\label{eq:Hamiltonian}
\end{gather}
where $\hat{x}^{\dagger}_{\mathbf{k}\sigma}$ creates an exciton  with 2D transverse momentum  $\mathbf{k}$, spin $\sigma$, and kinetic energy $\varepsilon_\mathbf k^x=k^2/2m_x$. Likewise, 
$\hat{c}^{\dagger}_{\mathbf{k}\sigma}$ creates  a photon  with momentum  $\mathbf{k}$, spin $\sigma$, and kinetic energy $\varepsilon_\mathbf k^c=k^2/2m_c+\delta$, where 
$\delta$ is the detuning. We have defined $\hat\rho_{\mathbf{q}\sigma}=\sum_{\mathbf k}\hat x_{\mathbf{k-q}\sigma}^\dagger\hat x_{\mathbf{k}\sigma}$ and  use units where the system volume
and $\hbar$  are  one. The first line of Eq.~\eqref{eq:Hamiltonian} describes excitons  coupled  
to  photons in the microcavity with Rabi frequency $\Omega$, giving rise to the formation of lower and upper exciton-polariton branches  with energies 
 $\varepsilon_\mathbf k=(\varepsilon_\mathbf k^c+\varepsilon_\mathbf k^x\pm\sqrt{\delta_{\mathbf k}^2+\Omega^2})/2$, where 
 $\delta_\mathbf k=\varepsilon_\mathbf k^c-\varepsilon_\mathbf k^x$. 

The second line in Eq.~\eqref{eq:Hamiltonian} describes the interaction between excitons with opposite and parallel spins with strengths $g_{\uparrow\downarrow}$ and $g_{\uparrow\uparrow}$ respectively. They  are both taken to be momentum independent, since their  typical length scale  is given by the exciton radius, which is much shorter than any other relevant length scale~\cite{Vladimirova2010}. The interaction between $\uparrow$ and $\downarrow$ excitons supports a bound state, the bi-exciton, and the corresponding scattering matrix ${\mathcal T}(p)$ therefore depends strongly on momentum and energy with a pole at the bi-exciton energy $\varepsilon_{\uparrow\downarrow}$.

As illustrated in Fig.~\ref{OverviewFig}(a), we consider a BEC of  exciton-polaritons with density $n_B$  
and spin-polarization $\uparrow$ created by a pump beam. A weaker probe beam creates a small density of exciton-polaritons with spin-polarization $\downarrow$, which can be regarded as impurities. Their interaction with the surrounding BEC leads to the formation of quasiparticles coined (Bose) polaron-polaritons~\cite{Bastarrachea2019,Levinsen2019}. In Fig.~\ref{OverviewFig} (b), the green line shows the energy of the lowest polaron-polariton branch as a function of the detuning $\delta$.  It is below the lower polariton energy 
$\varepsilon_\mathbf k=(\varepsilon_\mathbf k^c+\varepsilon_\mathbf k^x-\sqrt{\delta_{\mathbf k}^2+\Omega^2})/2$ shown by the red line in Fig.~\ref{OverviewFig}(b) due to interactions with the BEC of $\uparrow$ polaritons. To calculate the energy of the polaron-polariton branch in Fig.~\ref{OverviewFig} (b),  we have employed a diagrammatic approach where the quasiparticle energies 
appear  as poles in a $2\times2$ matrix Green's function $\mathcal G(p)=[\mathcal G^{-1}_0(p)-\Sigma(p)]^{-1}$ corresponding to the Hamiltonian in Eq.~\eqref{eq:Hamiltonian}.
 Here, $\mathcal G_0(p)=\text{diag}(\omega-\varepsilon_{\mathbf{p}}^{x},\omega-\varepsilon_{\mathbf{p}}^{c})$ is a diagonal matrix
 whose elements are the noninteracting exciton/photon Green's functions and the 
 light-matter coupling is described by the off-diagonal self-energies $\Sigma_{cx}=\Sigma_{xc}=\Omega/2$. Finally, the self-energy
 \begin{align} 
\Sigma_{xx}(p)=n_B\mathcal C_0^2{\mathcal T}(p),
\label{ExcitonSelf1}
\end{align}
describes the interaction of the $\downarrow$ exciton with the $\uparrow$ exciton BEC including the Feshbach resonance non-perturbatively.~\cite{Bastarrachea2019,SM}.
 $\mathcal C_{\mathbf q}^{2}=(1+\delta_{\mathbf q}/\sqrt{\delta_{\mathbf k}^2+\Omega^2})/2$ is 
the Hopfield coefficient of the polaritons in the BEC~\cite{Hopfield1958}.  In the calculation of the energies in  Fig.~\ref{OverviewFig} (b),  we use  realistic experimental parameters with  a Rabi splitting $\Omega=3.45 \text{meV}$, exciton mass $m_x\approx 0.16 m_{e}$ with  $m_e$ the electron mass, and  
$m_c=10^{-4}m_x$~\cite{Takemura2014,Ferrier2011,Ciuti1998,Rodriguez2016,Delteil2019}. The density of the BEC is $n_B=5\times 10^{10}\text{cm}^{-2},$  the direct exciton-exciton coupling  $g_{\uparrow\uparrow}\approx 3\mu\text{eV}\mu\text{m}^2 $, and the energy 
of the bi-exciton state is $\varepsilon_{\uparrow\downarrow}=0.7\text{meV}$.

 \emph{Effective interaction.--}
 Our focus here is on the interaction between polaron-polaritons  mediated by sound modes in the BEC. 
 We first calculate the mediated interaction between two  bare $\downarrow$ excitons using a non-perturbative approach that includes strong Feshbach correlations between a 
 pair of  $\uparrow\downarrow$ excitons  exactly, combined with a Bogoliubov theory generalised to the steady-state BEC at hand~\cite{Carusotto2004,Ciuti2005,Carusotto2013}. 
This gives
\begin{gather}\label{MediatedInteraction}
V(p,p'; q) =n_B\mathcal C_0^2\mathcal C_{\mathbf q}^2
\mathbf{T}^\intercal(p)\mathbf{G}^{\text{LP}}(q)\mathbf{T}(p')
\end{gather} 
for the mediated interaction between two $\downarrow$ excitons  with energy/momentum $p-q/2$ and $p'+q/2$ scattering into final states with energy-momentum 
 $p+q/2$ and $p'-q/2$.
Here $\mathbf{G}^{\text{LP}}(q)$ is a  $2\times2$ matrix containing the normal and anomalous Green's functions describing sound propagation in the  BEC~\cite{SM}.  
We have  defined the vector $\mathbf{T}^\intercal(p)=\begin{bmatrix}\mathcal{T}(p+q/2)&\mathcal{T}(p-q/2)\end{bmatrix}$ describing the 
scattering between the sound mode and the excitons, and the Hopfield factors in Eq.~\eqref{MediatedInteraction}  are  due to the fact that 
it is only the excitonic part of the BEC that scatters. The factor $n_B$ reflects that the interaction is mediated by the BEC. 
{For weak $\uparrow\downarrow$ interaction,   Eq.~\eqref{MediatedInteraction}
  takes the familiar form of $V(q)=n_B\mathcal C_0^2\mathcal C_{\mathbf q}^2g_{\uparrow\downarrow}^2\chi(q)$, which 
 simply describes an induced interaction mediated by sound waves with a strength determined by the density-density response function $\chi(q)$ of the BEC, 
 and with a range determined by the BEC  coherence length $\propto 1/\sqrt{2m_Bn_Bg_{\uparrow\uparrow}}$~\cite{Viverit2000}. } 
 For stronger $\uparrow\downarrow$ interaction close to the Feshbach resonance, the mediated interaction $V(p,p'; q)$ depends on both the incoming and the 
 transferred frequency/momenta due to the lack of Galilean invariance and the finite speed of sound in the BEC.

We can now calculate the effective interaction  between   two $\downarrow$ polaron-polaritons, which is the  physically relevant quantity. 
As in the derivation of Landau's Fermi liquid theory~\cite{BaymPethick1991book,NegeleOrland1998}, this is obtained by evaluating the mediated interaction on-shell 
between two  polaron-polaritons. 
Consider for concreteness the   scattering between two polaron-polaritons in the  branch shown by the green line in  Fig.~\ref{OverviewFig}(b).
Taking both the incoming and outgoing momenta to be zero, we obtain from  Eq.~\eqref{MediatedInteraction}
\begin{gather}
f(\mathbf 0,\mathbf 0)=-\frac{2\mathcal{T}^2(\mathbf 0,\varepsilon_0)}{3 g_{\uparrow\uparrow}}
Z_{0}^4,
\label{effectiveInt}
\end{gather}
where  $\varepsilon_{\mathbf p}$ and $Z_{\mathbf p}$ denote the energy and exciton residue of a  polaron-polariton with momentum ${\mathbf p}$.  The effective interaction in Eq.~\eqref{effectiveInt} depends on the $\downarrow\uparrow$ scattering matrix ${\mathcal T}$ squared, reflecting that the basic mechanism is the emission and subsequent absorption of a sound mode in the BEC. The $1/g_{\uparrow\uparrow}$ dependence shows  that  the interaction is stronger the more compressible the BEC. Compared to the mediated interaction between two  impurities  in a 
conventional BEC~\cite{Viverit2000,Heiselberg2000,Camacho2018b,Charalambous2019}, Eq.~\eqref{effectiveInt} contains an additional factor 
 $2/3$ originating from the non-equilibrium nature of the  BEC, as well as the residue $Z_{\mathbf p}$.

Figure \ref{OverviewFig}(c) shows numerical results for the effective interaction $f(\mathbf 0,\mathbf 0)$ between two  polaron-polaritons 
 as function of the detuning $\delta$. We see that the interaction is attractive and increases in strength with the detuning,  becoming of the order 
$f(\mathbf 0,\mathbf 0)\sim \Omega/n_B$ when $\delta\gtrsim\Omega/2$. Assuming a polaron-polariton density $n$ not too small compared to $n_B$, this 
corresponds to a  large mean-field shift $nf(\mathbf 0,\mathbf 0)$ of the order of $n/n_B\Omega$.
  The mechanism for this strong interaction is two-fold. First, the exciton component, which is the one that interacts, increases with increasing $\delta$. Second, the energy of the attractive lower polaron-polariton approaches that of the bi-exciton with increasing $\delta$ giving rise to a strong 
 $\uparrow\downarrow$ exciton scattering and thereby a large effective interaction.  
  
\emph{Dimer states.--}
The strong effective interaction between two $\downarrow$ polaron-polaritons  combined with its long range nature determined by the coherence length of the BEC, and the 2D geometry raises the intriguing possibility of  bound states consisting of two $\downarrow$  polaron-polaritons. 
{To capture the presence of such dimers, we have to  account for the strong correlations arising as a consequence of } the repeated scattering   between two $\downarrow$ excitons via the exchange of sound modes in the BEC. This is done via 
  the Bethe-Salpeter equation (BSE) for the scattering matrix  $\Gamma$ between a pair of $\downarrow$ excitons, which 
 in the ladder approximation reads~\cite{Fetter1971}
\begin{gather}
\Gamma({ k}_1,{ k}_2;{k}_1-{ k}_3)=V({ k}_1,{ k}_2;{ k}_1-{ k}_3)+\sum_{q}V({ k}_1,{ k}_2;q)\nonumber \\
\times G({ k}_1-q)G({ k}_2+q)\Gamma({ k}_1-q,{ k}_2+q;{ k}_1-q-{ k}_3).
\label{BetheS}
\end{gather}
Here, $G(k)$ is the Green's function of the $\downarrow$ excitons coupled to light and interacting with the BEC.

It is very complicated to solve the BSE taking into account the full  momentum/energy dependence of the interaction $V({ k}_1,{ k}_2;{ k}_3)$, and we  therefore make two approximations. 
First, we employ a pole expansion of $G(k)$, which corresponds to assuming that the $\downarrow$ excitons exclusively exist in the  polaron-polariton state when they are unbound. 
 Second, we neglect retardation effects by setting the frequency to zero in the mediated interaction $V({ k}_1,{ k}_2;{ k}_3)$. Even when both approximations are performed, the numerical calculation is still quite involved.  Details are given in the Sup.\ Mat.~\cite{SM}. 

The blue line in Fig.~\ref{OverviewFig}(b) shows the binding energy $E$ of the dimer relative to the polaron-polariton  obtained from solving $\Gamma^{-1}(E)=0$ for zero 
center of mass momentum. 
Remarkably, we see that the mediated interaction indeed is strong enough to bind two polaron polaritons into a dimer state beyond a critical detuning $\delta\gtrsim0.5\Omega$.
The binding energy of the dimer increases with 
$\delta$ reflecting the increasing attraction in agreement with Fig.~\ref{OverviewFig}(c), and it becomes comparable to the energy shift of the polaron-polariton with respect  to the 
bare polariton. Since the polaron-polaritons are partly photons these dimers in turn correspond to bound states of two photons. The existence of an 
mediated photon-photon interaction strong enough to support bound states is a main result of this work. 

Note that in a vacuum there is always a bound state for any attractive interaction between two particles~\cite{LandauBook}. The reason the  bound state only exists beyond a critical 
detuning in the present case is due to many-body effects, which alter the one particle dispersion into that of a polaron-polariton and make the mediated 
interaction non-local, i.e.\ depending on all momenta~\cite{Camacho2018}. 

\begin{figure}[!ht]
\includegraphics[width=0.99\columnwidth]{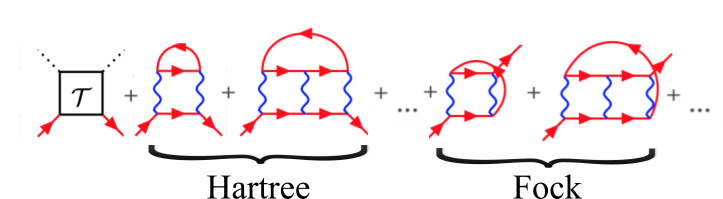}
\caption{The $\downarrow$ exciton self-energy with the first term describing  scattering with the BEC giving Eq.~\eqref{ExcitonSelf1} and the next two terms describing 
 the interaction with other   $\downarrow$ polaron-polaritons via  the mediated interaction giving Eq.~\eqref{ExcitonSelfenergy}. 
Red lines are the $\downarrow$ exciton propagator, dashed lines are $\uparrow$ polaritons  in the BEC, the box is the $\mathcal T$ matrix, and 
the wavy line is the mediated interaction in Eq.~\eqref{MediatedInteraction}. The first order term in the mediated interaction is included in the $\mathcal T$ matrix.
}
\label{FigFeynman}
\end{figure}

\emph{Light transmission.--} {We now show that the dimer state gives rise to a distinct line of transmitted light}. To do this, we  include the correlations leading to the formation 
of the dimer in the self-energy as 
\begin{align}
\Sigma_\text{in}(p)=nZ_{p}^2[\Gamma(p,0;0)+\Gamma(p,0;-p)].
\label{ExcitonSelfenergy}
\end{align}
{Equation \eqref{ExcitonSelfenergy} gives the energy shift of the  polaron-polariton due its interactions with other polaron-polaritons of density $n$ in the Hartree-Fock approximation.
We have replaced the induced interaction with the scattering matrix $\Gamma$ obtained from the BSE to include  strong coupling effects~\cite{Fetter1971,SM}.
The Feynman diagrams corresponding to Eq.~\eqref{ExcitonSelfenergy} are shown in Fig.~\ref{FigFeynman}.}
Adding $\Sigma_\text{in}(p)$ to Eq.~\eqref{ExcitonSelf1}  includes dimer formation in our many-body theory for the polaron-polaritons.  

In Fig.~\ref{OverviewFig} (d)-(f), we plot the $\downarrow$ photon  spectral function as a function of the detuning $\delta$. 
There are several polaron-polariton branches typical of the interplay between strong interactions and light 
coupling~\cite{Sidler2016,Levinsen2019,Bastarrachea2019,Tan2020}. The lowest branch corresponds to the green line in Fig.~\ref{OverviewFig}(b).
  Compared to the spectrum for vanishing polaron-polariton density shown in Fig.~\ref{OverviewFig} (d), the spectra for the polaron-polariton densities $n=0.075n_B$ (e) and $n=0.15n_B$ (f)
     have one 
  qualitative new feature: there is a new dimer  line, which comes  from the fact that when the energy of the incoming photon and a polaron-polariton already present 
  equals that of a dimer state,  
a bound state involving two photons is formed, which propagates through the BEC giving rise to light transmission. Comparing Fig.~\ref{OverviewFig} (e) and (f), 
we see that the strength of this dimer line increases with $n$  because  there 
are then more polaron-polaritons in the BEC to form dimers with.

\begin{figure}[h!]
\includegraphics[width=0.99\columnwidth]{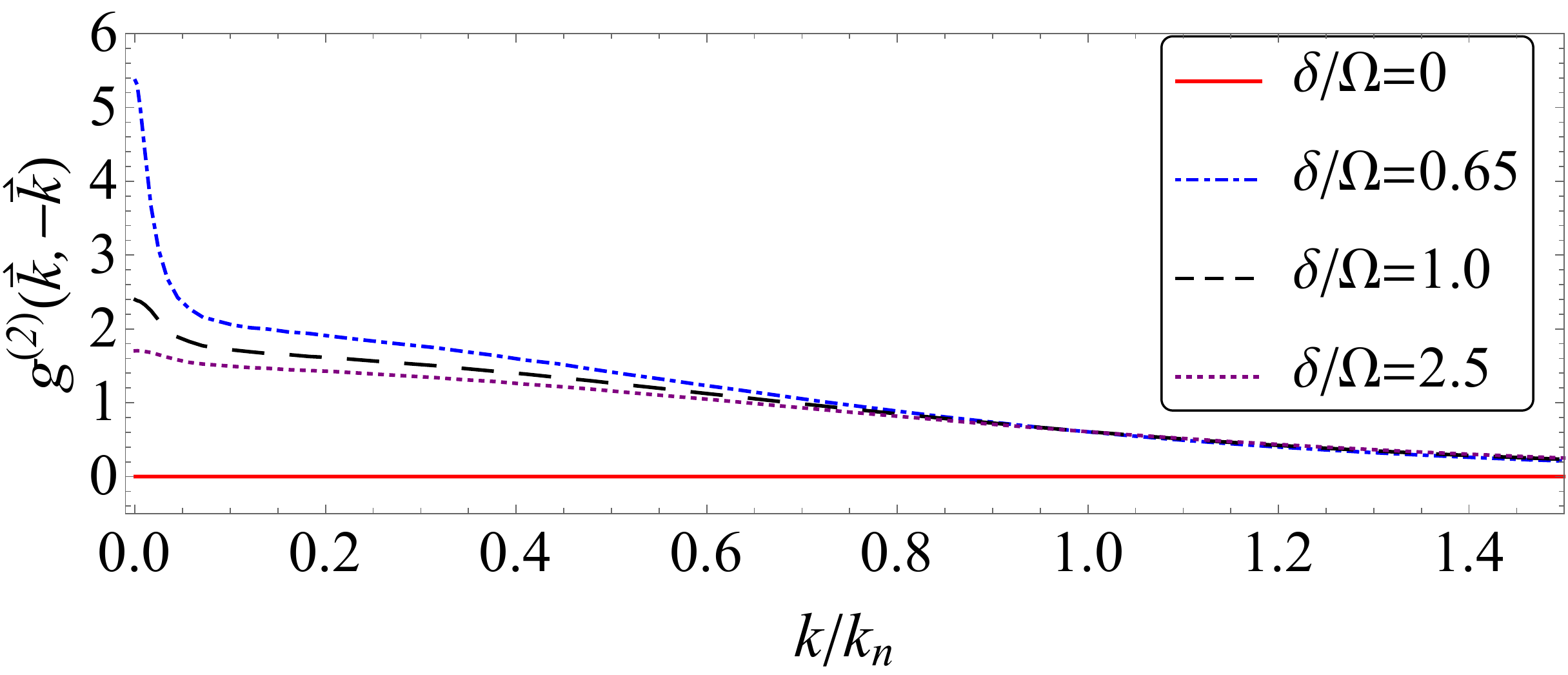}
\caption{(Color online)  Correlations between transmitted photons with opposite transverse momenta as parametrised by 
$g_{2}(\mathbf p,-\mathbf p)$ for different detunings. 
}
\label{FigCorrelations}
\end{figure}

\emph{Photon correlations.--}
We finally show that the wave function of the dimer is  imprinted 
in the transmitted light correlations. Figure \ref{FigCorrelations} plots the correlation function  $g_{2}(\mathbf p,-\mathbf p)=\langle a^\dagger_{\mathbf p} a_{\mathbf p}a^\dagger_{-\mathbf p}a_{-\mathbf p} \rangle-\langle a^\dagger_{\mathbf p} a_{\mathbf p}\rangle\langle a^\dagger_{-\mathbf p} a_{-\mathbf p}\rangle$ for 
 different detunings $\delta$, where, $a^\dagger_{\mathbf p}$ creates a
polaron-polariton with momentum ${\mathbf p}$. It is calculated using the wave function $|\Phi\rangle=\sum_{\mathbf p>0}\phi(\mathbf p)a^\dagger_{\mathbf p}a^\dagger_{-\mathbf p}|0\rangle$ of the dimer as $g_{2}(\mathbf p,-\mathbf p)=|\phi(\mathbf p)|^2-|\phi(\mathbf p)|^4$, where $|0\rangle$ is the vacuum state. Here,  $\phi(\mathbf p)$ is
 obtained from the BSE by mapping it onto an effective Schr\"odinger equation~\cite{SM,Camacho2018}. We see that $g_2$ indeed is non-zero when  there is a bound state. 
 The correlations moreover extend to increasingly high momenta with increasing $\delta$, which directly reflects the decreasing spatial size of the dimer wave function with 
 increasing binding energy. Figure \ref{FigCorrelations} therefore 
 demonstrates that  the dimer wave function can be measured directly from the 
   correlations of the transmitted light. 

\emph{Discussion.--}
Since we are considering a strongly correlated hybrid light-matter system, it is worth discussing our approach. First, the ladder approximation describing the formation of polaron-polaritons is surprisingly accurate for the analogous problem of atomic polaron formation~\cite{Tempere2009,Massignan2014,Jorgensen2016,Hu2016,Yan2019,Rath2013,Ardila2019,Lampo2017,Ardila2020}. An similar theory  for the mediated interaction between impurities  and dimer formation in an atomic BEC  has moreover been shown to be  remarkably accurate  even for strong interactions when benchmarked against Monte-Carlo calculations~\cite{Camacho2018b,Camacho2018}. Non-equilibrium Bogoliubov theory is known
 to be a reliable  description of the  polariton BEC~\cite{Carusotto2013}. 
 Since our results are  based on the existence of a linear sound spectrum in the medium, we expect them to be robust towards 
fragmentation of the BEC~\cite{Estrecho2018,Pieczarka2020}.
 The intrinsic decay due to photon leakage out of the cavity will not significantly affect our results, as long as the 
 resulting line widths are small compared to their separation. 
   Finally, our  approach is  based on the  well established microscopic 
 foundation  of Landau's theory of effective interactions between quasiparticles~\cite{NegeleOrland1998}.

Recently, it has been shown that the experimental findings in Refs.~\cite{Takemura2014,Takemura2014b,Takemura2017,Navadeh2019} are consistent with a fast decay  of 
the $\uparrow\downarrow$ bi-exciton underlying the Feshbach resonance~\cite{Bastarrachea2019}. Such a decay   will likely 
 decrease the strength of the mediated interaction. In order to see the dimers discussed here, one therefore needs clean samples.

To observe  the mediated strong photon-photon interaction and the dimer state,  one could  
use  a spectrally broad probe beam creating both the dimer states and the polaron-polaritons from which they are formed.
The intensity of the dimer transmission line will scale as the intensity 
of the probe beam squared clearly reflecting its non-linear nature. 
Note that one cannot   include the effective photon-photon interaction straightforwardly
in a  non-linear equation for the light fields, since it is a strong coupling many-body phenomenon.

We briefly comment on the relation of our results to those in Ref.~\cite{Bastarrachea2019}. Reference \cite{Bastarrachea2019} considered the propagation of a 
single $\downarrow$ polaron-polariton in a $\uparrow$ BEC. This corresponds to the  $n=0$ limit of the present paper where there are no dimers. 
The long-range effective interaction between two $\downarrow$ polaron-polaritons 
considered here is  a genuine many-body effect only present for $n>0$, where it leads to  non-linear effects such as dimer formation.

\emph{Outlook.--} 
We have shown how the effective interaction between polaron-polaritons in a semi-conductor can be strong enough to support 
 dimer states involving two photons. This gives rise to a new transmission line where the wave function is imprinted directly in the correlations of the transmitted light. 
Our results  demonstrate how hybrid light-matter systems
offer powerful new ways to probe many-body physics, in this case  effective interactions  
which are a key ingredient in Landau's quasiparticle theory. 
 The possibility to engineer mediated strong photon-photon interactions  in a semiconductor microcavity moreover 
  opens the door to realising highly non-linear optics in a solid state setting and engineering scalable optoelectronic devices such as  gates, switches, and transistors~\cite{Carusotto2013,Sanvitto2016,Liew2008,Shelykh2010,Amo2010,Gao2012,Nguyen2013,Espinosa2013,Zasedatelev2019,Dreismann2016,Ballarini2013}. 
  We finally note that strongly interacting 
   polaron-polaritons can also be realised in atomic systems~\cite{Grusdt2016,Camacho2020,Nielsen2020}.
  
\begin{acknowledgments}
We acknowledge financial support from
 the Villum Foundation and the Independent Research Fund Denmark - Natural Sciences via Grant No. DFF - 8021- 00233B.

\end{acknowledgments}

\bibliography{Polariton}

\end{document}